\documentclass[10pt]{iopart}
\usepackage{graphicx}

\usepackage{float}
\usepackage[english]{babel}
\usepackage[english]{cleveref}
\usepackage{csquotes}
\usepackage[inline]{enumitem}

\usepackage[dvipsnames]{xcolor}         
\usepackage{booktabs}                   
\usepackage{cite}
\usepackage{derivative}
\usepackage{xfrac}

\usepackage[per-mode=symbol-or-fraction, uncertainty-mode=separate]{siunitx}

\usepackage{xspace}

\begin{document}

\title{Response of a Cold-Electron Bolometer in a coplanar antenna system}

\author{D~A~Pimanov$^1$, A~L~Pankratov$^{1,2}$, A~V~Gordeeva$^1$, A~V~Chiginev$^{1,2}$, A~V~Blagodatkin$^{1,2}$, L~S~Revin$^{1,2}$, S~A~Razov$^{1}$,  V~Yu~Safonova$^{1,2}$, I~A~Fedotov$^2$, E~V~Skorokhodov$^2$, A~N~Orlova$^2$, D~A~Tatarsky$^2$, N~S~Gusev$^2$, I~V~Trofimov$^3$, A~M~Mumlyakov$^3$, M~A~Tarkhov$^3$}
\address{$^1${Nizhny Novgorod State Technical University n.~a. R.~E.~Alekseev}, {Nizhny Novgorod}, {Russia}}
\address{$^2${Institute for Physics of Microstructures of RAS}, {Nizhny Novgorod}, {Russia}}
\address{$^3${Institute of Nanotechnology of Microelectronics of RAS}, {Moscow}, {Russia}}
\eads{\mailto{alp@ipmras.ru}, \mailto{rls@ipmras.ru}}

\begin{abstract}
Cold electron bolometers have shown their suitability for use in modern fundamental physical experiments. Fabrication and measurements of the samples with cold-electron bolometers integrated into coplanar antennas are performed in this study. The bolometric layer was made using combined aluminum-hafnium technology to improve quality of aluminum oxide layer and decrease the leakage current. The samples of two types were measured in a dilution cryostat at various temperatures from 20 to 300 mK. The first sample with Ti/Au/Pd antenna shows response in the two frequency bands, at 7--9 GHz with bandwidth of about 20\%, and also at 14 GHz with 10\% bandwidth. The NEP below 10 aW/$\sqrt{\rm Hz}$ is reached at 300 mK for 7.7 GHz signal. The second sample with aluminum made antenna shows response in the frequency range 0.5--3 GHz due to the effect of kinetic inductance of superconducting aluminum.
\end{abstract}

\vspace{2pc}

\submitto{\SUST}

\maketitle

\ioptwocol

\section{Introduction}

Modern problems of fundamental physics, such as search for circular B-mode of Cosmic Microwave Background radiation \cite{ABS,Sci,Lam,Lite,QB,Milli} and Dark Matter search experiments \cite{Lamore,QUAX,Sikivie,Sushkov} lead to quite tough sensitivity requirements of quantum sensors. 

The Cold Electron Bolometers (CEBs) \cite{Kuzmin_2002,Kuzmin_2012}, utilizing direct electron cooling of a nanoabsorber \cite{ElCool,ElCool2}, have proven to be highly sensitive detectors \cite{SWith}, demonstrating photon noise sensitivity level \cite{Brien_2014,Brien_2016,Gord_2017,Olimpo}. Thanks to the electron cooling effect \cite{Mart_93,Mart_94,Anghel_2001,Rajauria,Vas_2009,ONeil,Vas_2012,Vas_2013,ONeil2,Nguyen,Nguyen2} and tiny volume of a nanoabsorber, CEBs are suitable for space applications, since they can be operated in $^3$He sorption fridges \cite{Bhatia} and have record cosmic rays immunity \cite{Salatino_2014}.

While initially detectors with electron cooling were suggested for X-ray frequency range \cite{Mart_95},  recently it has been predicted that CEBs can be used as single photon detectors with photon wavelengths up to 1 cm \cite{ph_det_CEB}. To this end, it is interesting to design and fabricate CEB samples, embedded into coplanar antennas, intended to receive signals in GHz frequency range, which was not implemented before. 

Here we present first designs and tests of a cold electron bolometer embedded into a coplanar antenna. The modifications of an absorber layers using combined aluminum/hafnium technology in order to decrease leakage currents are discussed. This research is mostly aimed at temperatures of 250 and 300 mK, that are typical for single and double stage sorption $^3$He fridges. The response to microwave signal is investigated and two working frequency bands of 7-9 GHz and 14 GHz are revealed. It is demonstrated that the designed receiver can reach NEP below 10 aW/$\sqrt{\rm Hz}$ at 300 mK.

\section{Design and fabrication}

\begin{figure}[h]
    \includegraphics[width=\linewidth, keepaspectratio]{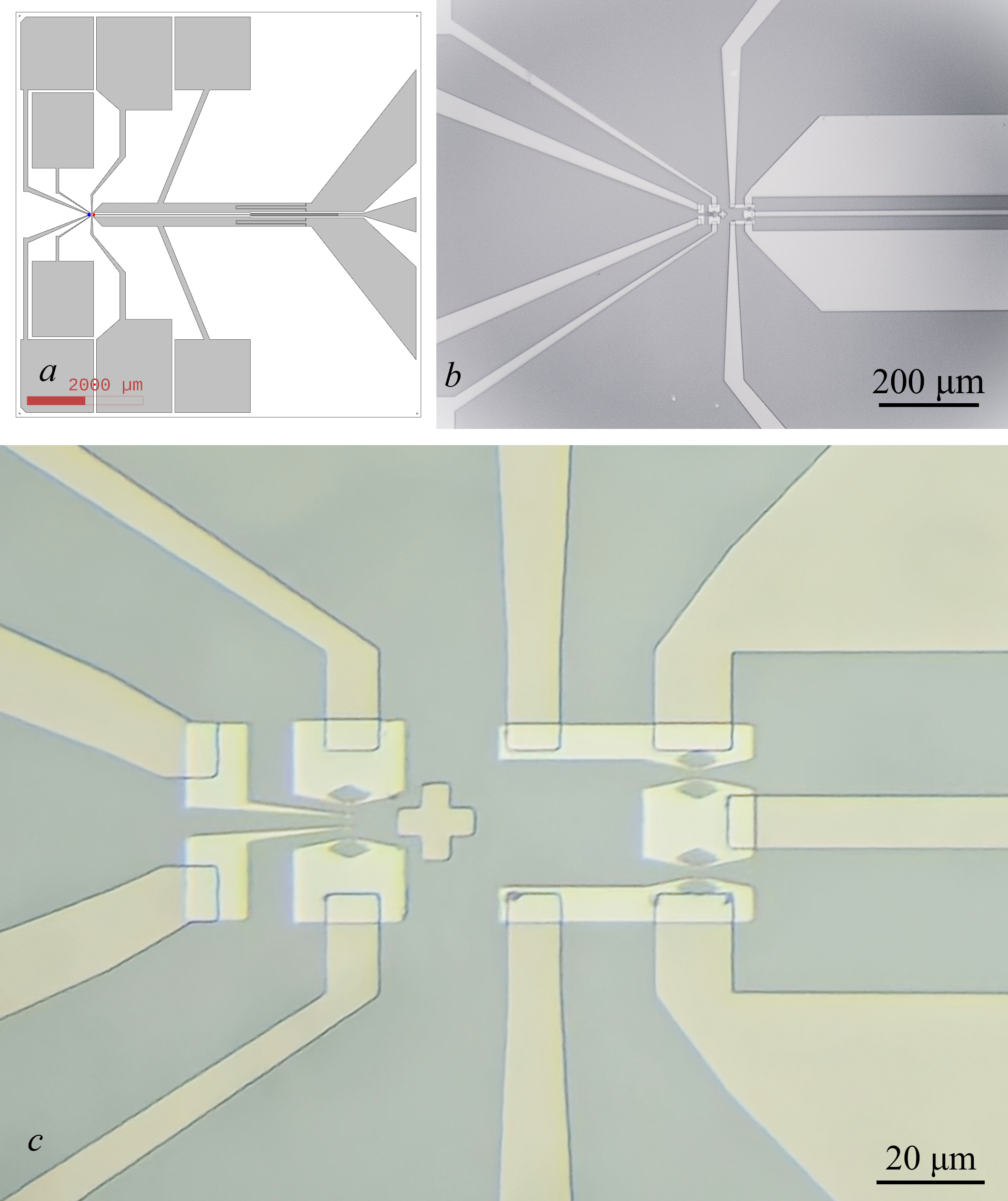} 
    \caption{({\it a}) The layout of a chip with two CEBs embedded into a coplanar antenna (right, red color) and the test structure (left, blue color). ({\it b}) Optical microscope image of a chip with two CEBs embedded into a coplanar antenna (right) and the test structure (left). ({\it c}) An enhanced image of a chip with two CEBs embedded into a coplanar antenna (right) and the test structure (left) taken from a metallographic microscope.}
    \label{fig:sample_design1}
\end{figure}
The scheme and images of a sample of the receiving system, consisting of two CEBs, embedded into a coplanar antenna, and a test structure, are presented in Fig. \ref{fig:sample_design1}. The CEBs, representing double superconductor-insulator-normal metal (SIN) junctions, were made using optical lithography and electron beam lithography. The deposition of materials was performed by electron beam evaporator using self-aligned shadow evaporation technique \cite{ElCool}.

First, antennas and contact pads based on a Si substrate were fabricated using laser writer and electron beam evaporator. This layer consists of 10 nm of titanium, 100 nm of gold and 20 nm of palladium.

For the fabrication of CEBs with 100 nm absorber several resists with different sensitivities are used. This allows to form an undercut needed for shadow evaporation technique. It is suitable to use methyl methacrylate (MMA) and AR 600 in our technology process since AR 600 is chemically uncoupled from MMA. N-Amyle Acetate and MIBK:IPA 1:1 developers are used for top and bottom resists, respectively. The exposure was performed in an electron microscope with a lithographic facility and a laser interference table.  

After the resistive mask was formed, the following films were deposited by electron beam evaporation. First, a Hf film with 10 nm thickness, then a Fe film with 1.2 nm thickness and after that Al film with 5.3 nm thickness were deposited along the normal to the sample surface. Further, oxidation was carried out in the oxygen for 5 minutes at 10 Torr pressure. After the formation of the oxide layer, 60 nm  and 70 nm thick aluminum films were deposited, at angles of +45 and -45 degrees to the normal of the sample. Finally, the lift-off processing was performed. During this process, the unexposed resist, and along with it, excess materials from unnecessary places, were removed. As a result, SIN tunnel junctions, embedded into an antenna, were fabricated.

It should be noted that the normal absorber was fabricated using superconducting/ferromagnetic bilayer to suppress superconductivity and also Andreev heating current in the absorber to improve electron cooling \cite{Vas_2013, ElCool}. In addition, Hf layer has been deposited as the first layer, underneath of Al/Fe bilayer. First, Hf has reduced electron-phonon coupling, by a factor of three lower than aluminum. This should potentially increase responsivity of a detector and decrease a heat leak from a substrate to a nanoabsorber. Second, Hf has higher density, thus aluminum, deposited above Hf layer, should have higher uniformity. Further oxidation of such aluminum layer should lead to more uniform, thinner and dense oxide layer thus reducing leakage current. 

\begin{figure}
 \includegraphics[width=\linewidth, keepaspectratio]{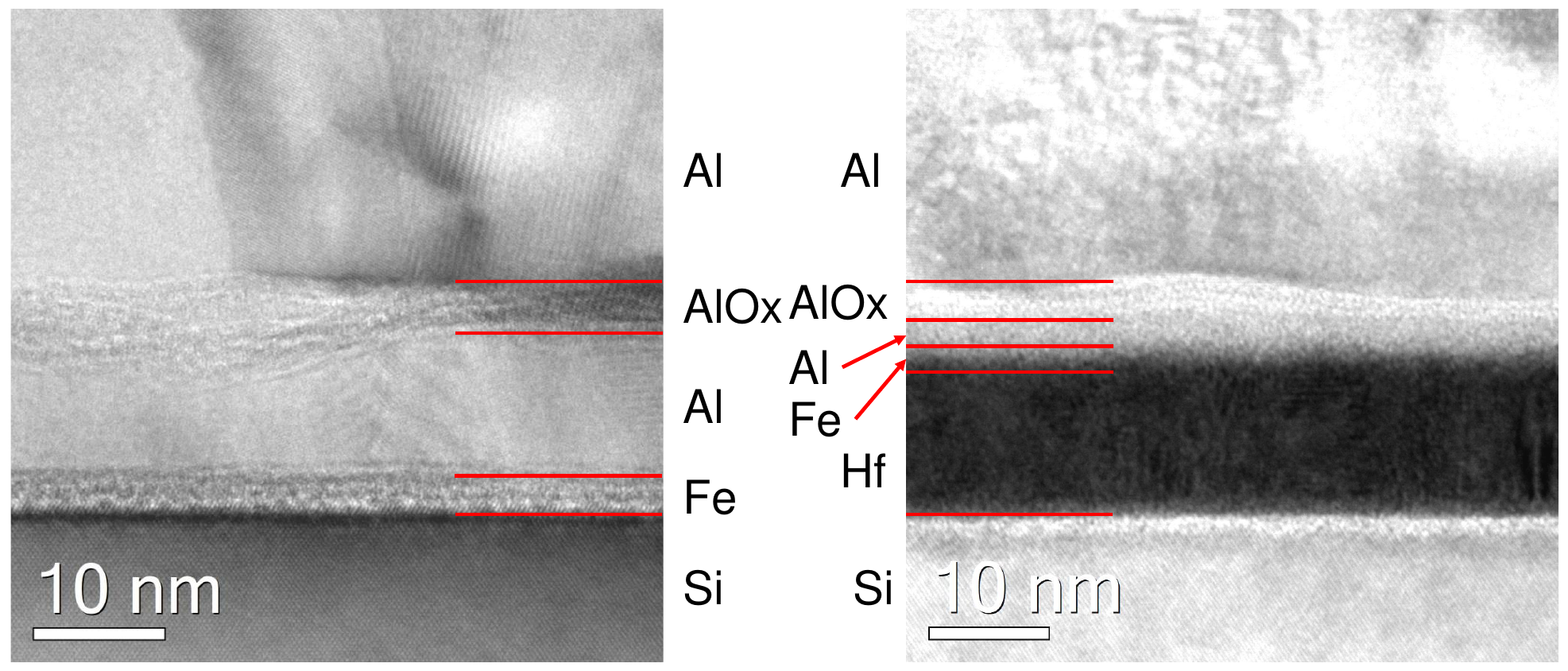}    
    \caption{Bright field TEM micrograph of cross-section of sample layers using the standard technology for manufacturing of cold electron bolometers (left) and the modification of this technology using hafnium (right).}
    \label{fig:sample_layers}
\end{figure}

To test these assumptions, similar samples using a standard technology with Al/Fe layer \cite{Olimpo,ElCool} and an improved Al/Fe/Hf technology were compared using a conventional Transmission Electron Microscopy of samples cross-sections (TEM), see Fig. \ref{fig:sample_layers}. Indeed, one can see that in the latter case the tunnel barrier has become more uniform and thinner due to the higher density and uniformity of the hafnium sublayer. Further, such CEB samples with Al/Fe/Hf absorber were measured in a dilution fridge and investigated under irradiation of a microwave signal.

In addition, a few samples have been manufactured using  typical technology for fabrication of superconducting  qubits \cite{Moskalev_2023}. Here, the first layer, containing DC lines, contact pads and the coplanar antenna, is made entirely of aluminum. To provide better adhesive properties of this aluminum layer to the silicon substrate, 5 nm titanium sublayer is deposited underneath. The design of the sample is completely the same, so the materials used for the first layer is the only difference. Before deposition of the bolometric layer, the natural aluminum oxide film from antenna layer was removed by argon ion etching, which is a complex procedure, not always leading to the desirable results. 

\section{Measurements}
\begin{figure}
 \includegraphics[width=\linewidth, keepaspectratio]{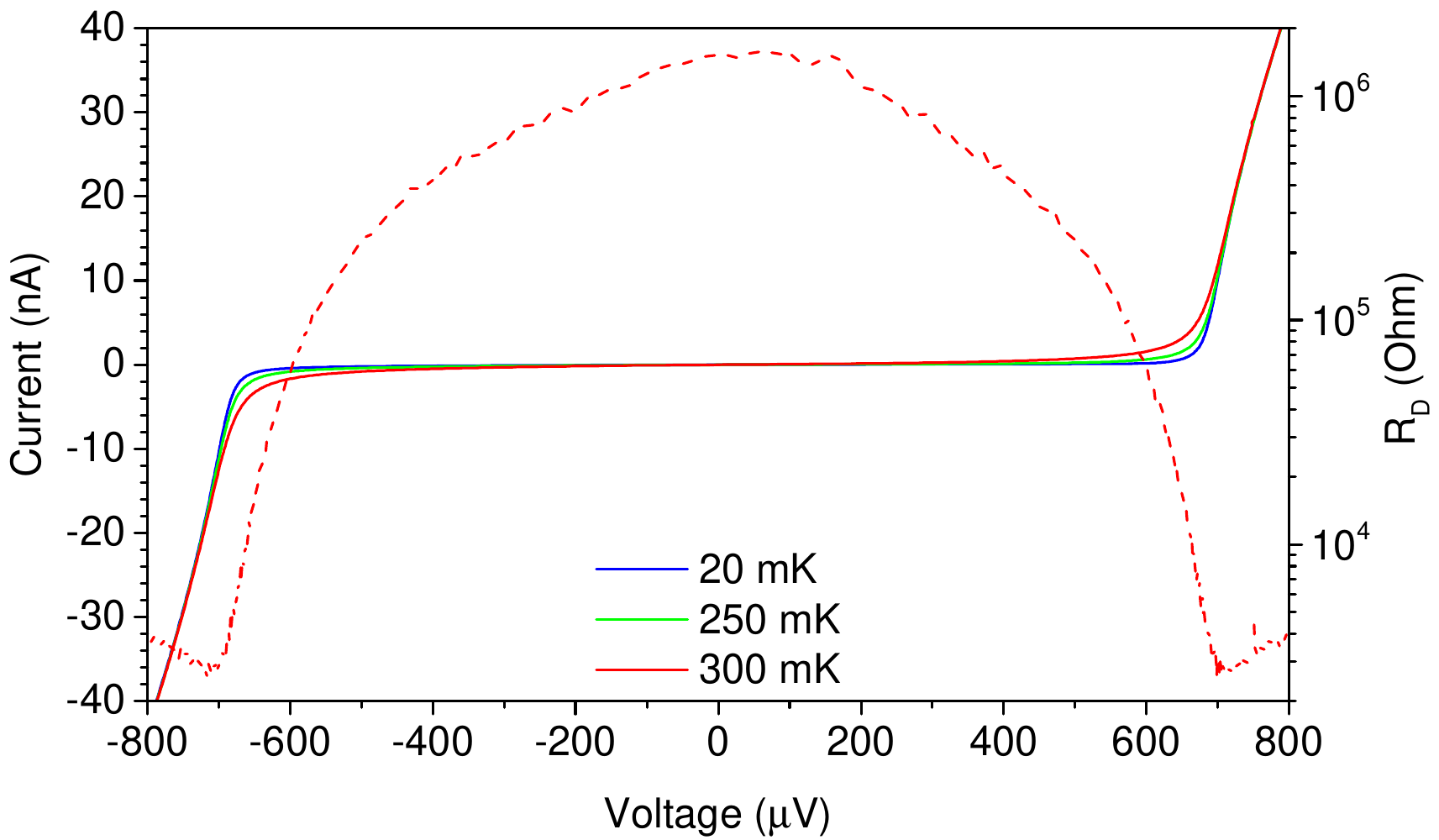}    
    \caption{Current-voltage characteristics of a bolometric structure measured at temperatures of 20, 250 and 300 mK. Dashed curve shows differential resistance of the structure, measured at the temperature of 300 mK.}
    \label{fig:sample_IV}
\end{figure}
Figure \ref{fig:sample_IV} shows the current-voltage characteristics (IVCs) of the sample, which consists of two cold electron bolometers connected in series in dc current and in parallel in microwave signal, see Fig. \ref{fig:sample_design1}c. The IVCs were measured at temperatures of 20 (blue), 250 (green) and 300 (red) mK in a dilution refrigerator without an external signal (solid curves). One can see that the IVC curves in Fig. \ref{fig:sample_IV} are rather steep even for 300 mK temperature, that signals about good sample quality and low leakage current. This is confirmed by the differential resistance plot (red dashed curve) for the temperature of 300 mK. Computing the ratio of differential resistance at zero bias current (maximal value) to the normal resistance, one gets the value about 250, while for the temperature of 20 mK this ratio exceeds 1000.

\begin{figure}
 \includegraphics[width=\linewidth, keepaspectratio]{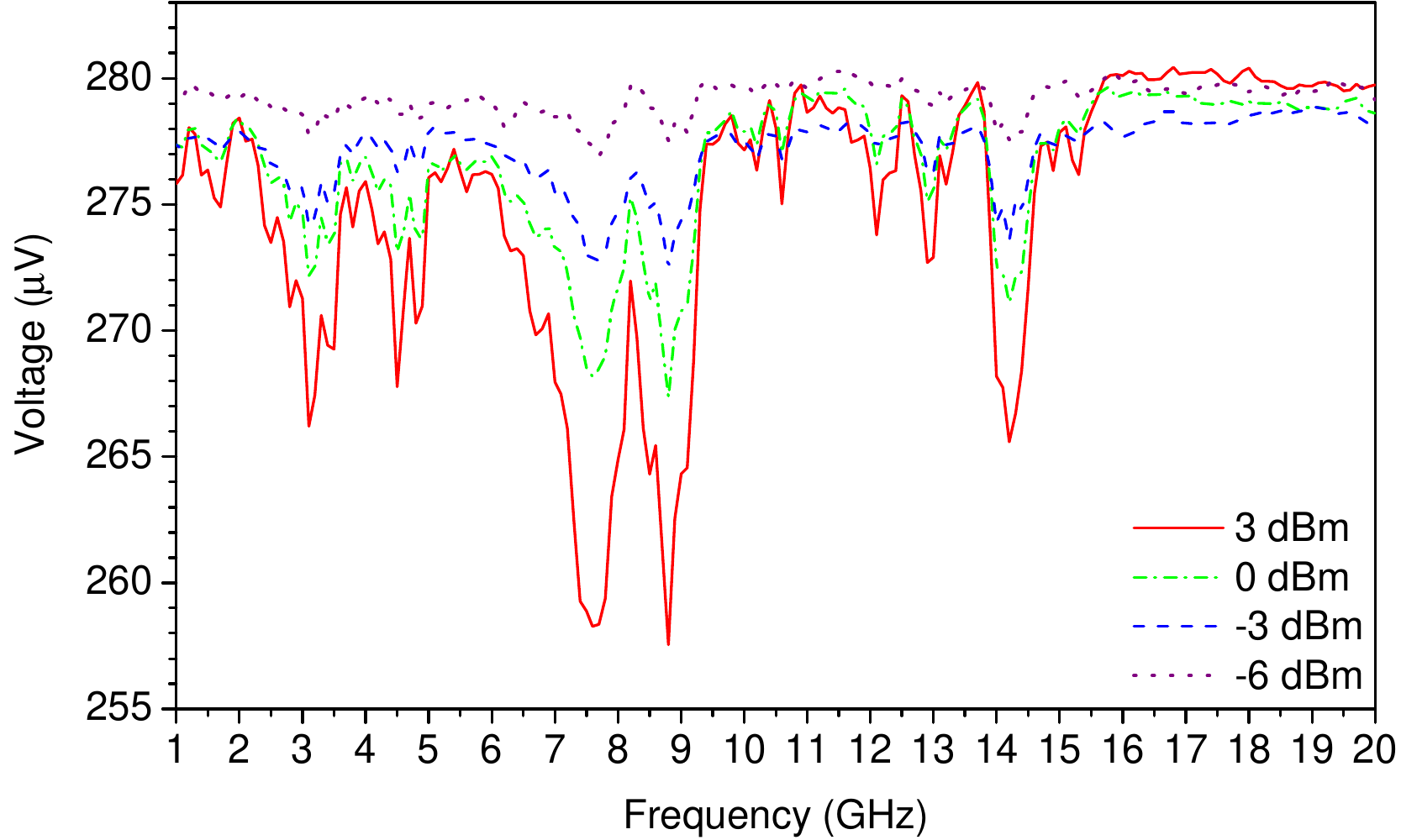}    
    \caption{Amplitude-frequency characteristics of the sample measured at the temperature of 300 mK.}
    \label{fig:sample_response}
\end{figure}
The amplitude-frequency characteristics (AFC) in a low power range were measured varying the synthesizer frequency from 1 GHz to 20 GHz, see Fig. \ref{fig:sample_response}.
The signal from the synthesizer was fed to the sample, located at the dilution chamber plate of the cryostat, via cryogenic coaxial cables with additional coaxial attenuators, which provide the total attenuation of 65 dB to suppress background radiation from room temperature environment. The measurements were carried out for the bias current leading to the maximal voltage response of the bolometer, which was set to 350 pA. To find this optimal point, it is necessary to plot the voltage difference on the current-voltage curves with and without a signal at the same current. The response curves for the power values at the synthesizer output of -6, -3, 0 and 3 dBm can be seen in Fig. \ref{fig:sample_response}. Here, two frequency ranges with the maximal response can be selected. The first one is 14 GHz, which is the main frequency range of the antenna, as was intended in the electromagnetic modelling of the design. However, in this frequency range the response is somewhat weaker than at the double minimum of 7.7 and 8.8 GHz due to significantly higher losses about 6 dB of the used coaxial line at 14 GHz. This coaxial line will be improved by change of coaxial attenuators. The detector resonances are rather narrow, about 10\% at 14 GHz and about 20\% at 7-9 GHz.
The maximum value of the sample response at the synthesizer power of 3 dBm is about 23 \textmu V.

\begin{figure}[h]
 \includegraphics[width=\linewidth, keepaspectratio]{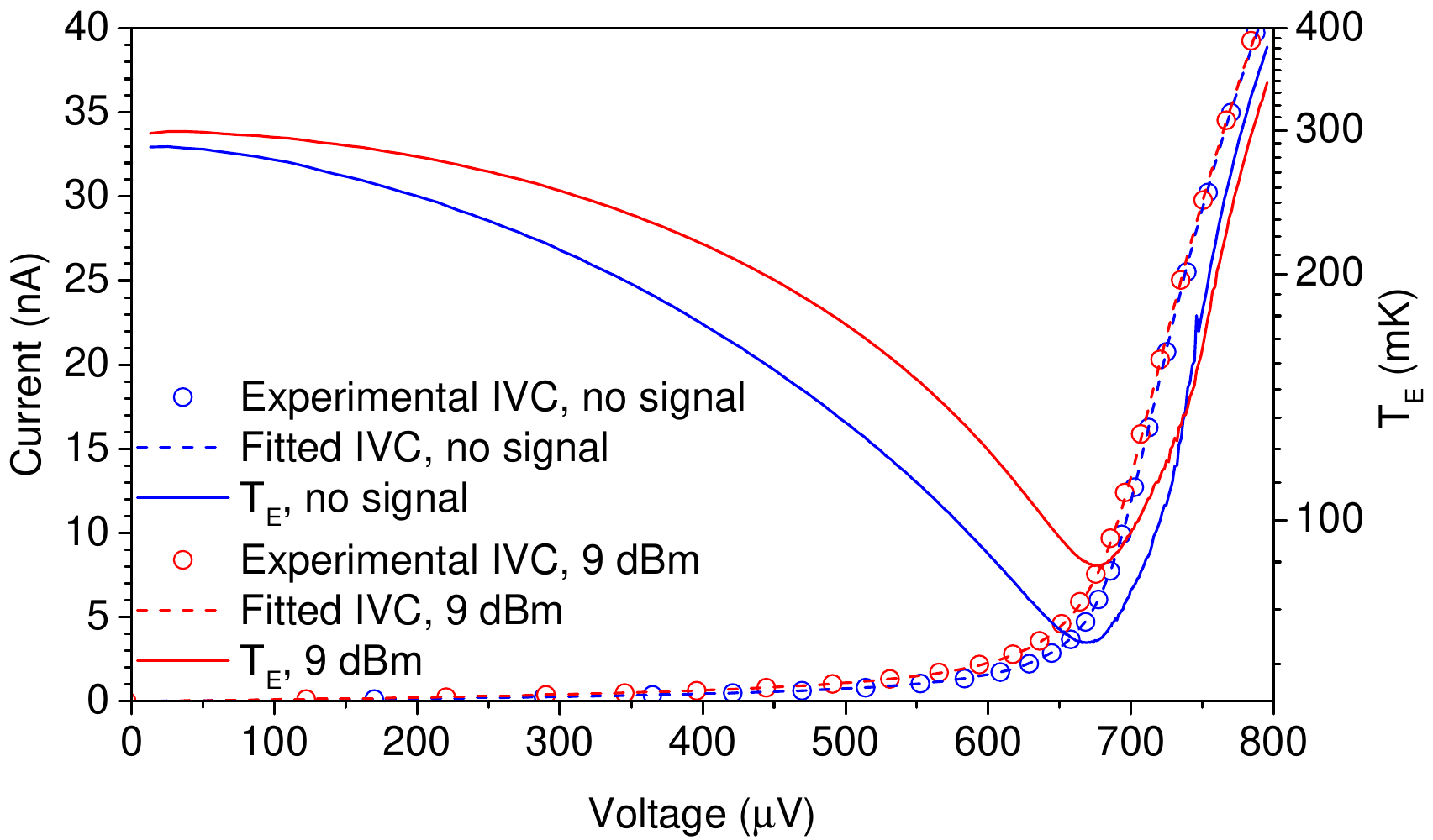}    
    \caption{The current-voltage characteristics (circles) of the sample, measured at 300 mK without (blue) and with the signal at 7.7 GHz (red). Fittings of the experimental curves using heat balance equations are shown by dashed curves. Electron temperature is shown by solid curves. }
    \label{fig:sample_Te}
\end{figure} 
To find the detected power, and also responsivity and NEP of the detector, the current-voltage curves were fitted using the heat balance equations,  described in detail in \cite{Golubev,ElCool2}, without and with 7.7 GHz signal, see Fig. \ref{fig:sample_Te} for the temperature of 300 mK. In addition to the heat balance equations, the electronic temperature can be calculated from the equation for the quasiparticle current \cite{Golubev}, which gives correct results, since in our designs the Andreev current and, respectively, the zero bias current peak, are suppressed due to the use of hybrid Al/Fe absorber \cite{ElCool}. In the absence of a power load, the considered sample demonstrates efficient electron cooling from 300 mK to 71 mK, see blue solid curve in Fig. \ref{fig:sample_Te}. Even at the high power load of 9 dBm at the synthesizer output, the electron cooling is still working effectively, with minimal electron temperature below 100 mK. 

\begin{figure}
 \includegraphics[width=\linewidth, keepaspectratio]{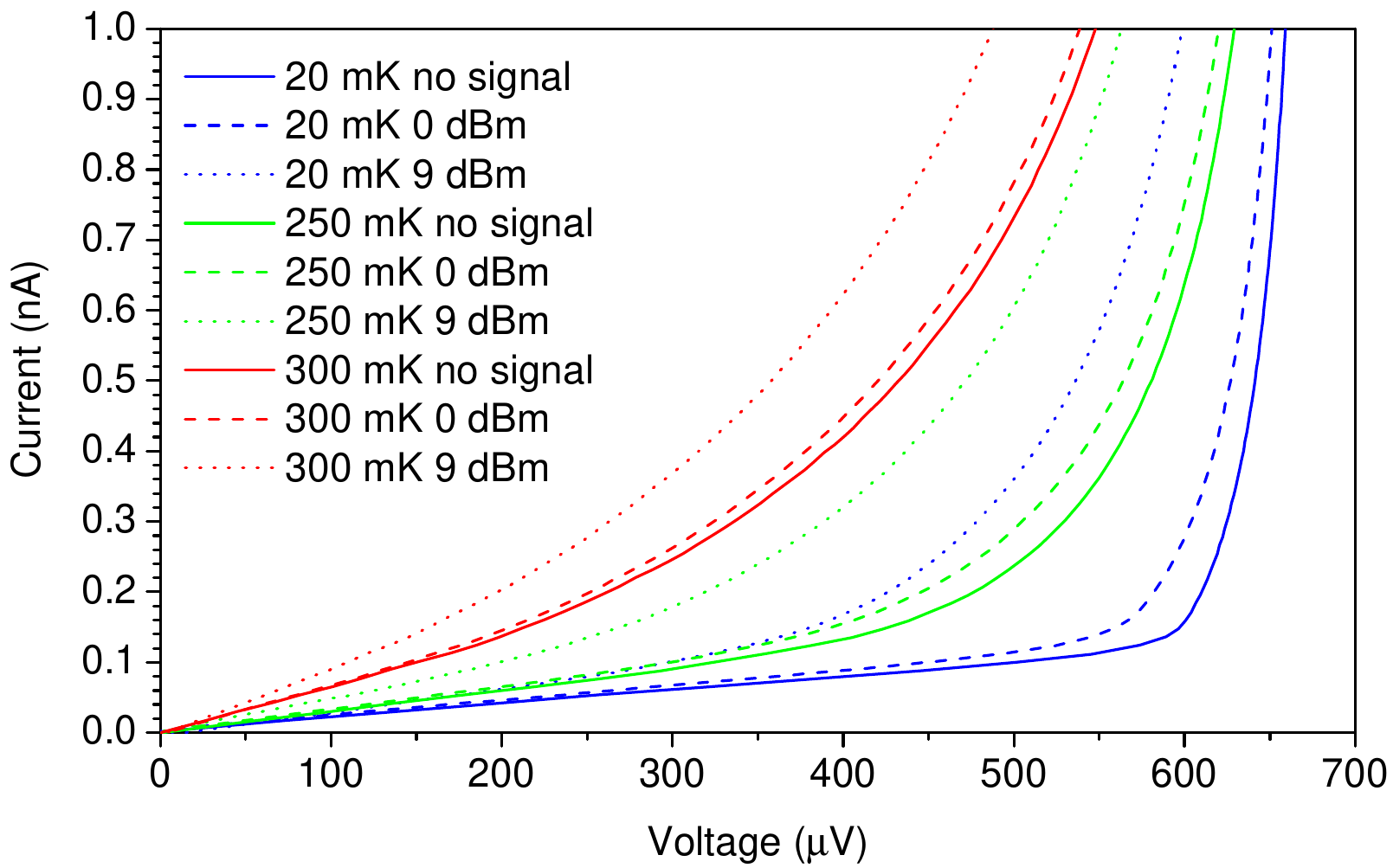}    
    \caption{The current-voltage characteristics of the bolometric structure measured at temperatures of 20, 250 and 300 mK with and without microwave radiation with different microwave powers at the synthesizer at frequency of 7.7 GHz.}
    \label{fig:sample_IV_RF}
\end{figure}
In Fig. \ref{fig:sample_IV_RF} the current-voltage characteristics for various temperatures and power loads are shown in more detail for currents below 1 nA.
The dashed curves in Figure \ref{fig:sample_IV_RF} were measured at the power of 0 dBm at the synthesizer frequency of 7.7 GHz and the dotted curves -- at the power of 9 dBm. 
Solid curves were obtained without an external signal.
According to the fitting of the measured current-voltage characteristics at 300 mK, the power, received by the bolometer at 0 dBm synthesizer power is about 10 fW, and at 9 dBm is about 80 fW. It is seen that with decreasing temperature, the response at the same synthesizer power level increases.   

\begin{figure}[h]
 \includegraphics[width=\linewidth, keepaspectratio]{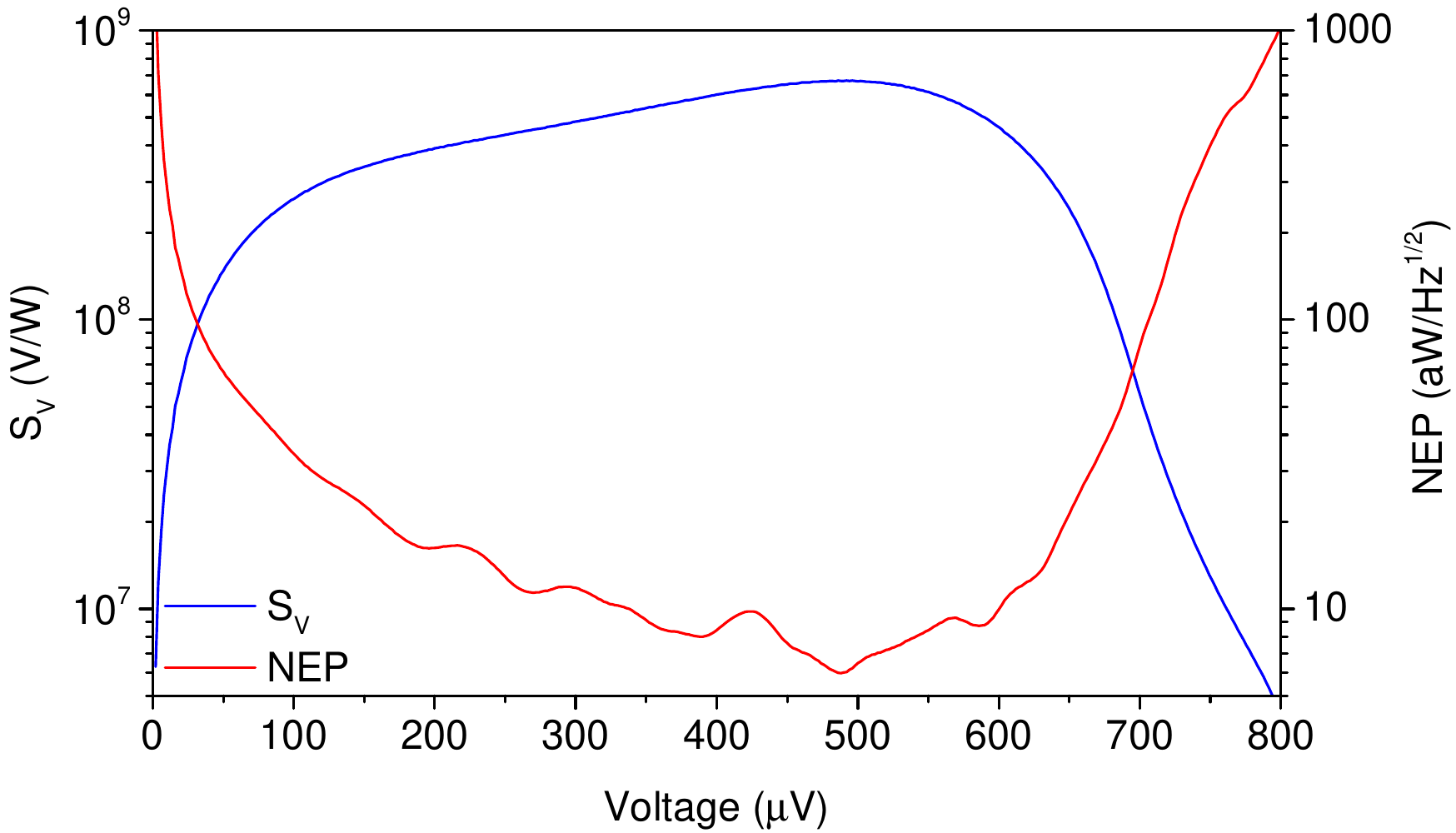}    
    \caption{Responsivity (blue curve) and NEP (red curve) of the sample at 300 mK temperature.}
    \label{fig:sample_Si_NEP}
\end{figure}
From the performed fitting of the received power in the weak signal limit at 300 mK temperature, one can get the maximal responsivity to be equal $6.7\times10^{8} \, \rm{V/W}$, see the blue curve in Fig. \ref{fig:sample_Si_NEP}. Correspondingly, the required Noise Equivalent Power (NEP) can be calculated as the total bolometer noise divided by the responsivity, which is shown by the red curve in Fig.  \ref{fig:sample_Si_NEP}. While the noise of the considered sample is limited by the noise of room temperature amplifier AD745 by 5 nV/$\sqrt{\rm Hz}$ level, the minimum NEP value of the sample, operating at 300 mK, is rather low, about $6 \, \rm{aW/\sqrt{Hz}}$. This is only twice higher than the internal NEP of a single bolometers of a similar design \cite{Olimpo}.  

\section{Sample with aluminum antenna}

\begin{figure}
 \includegraphics[width=\linewidth, keepaspectratio]{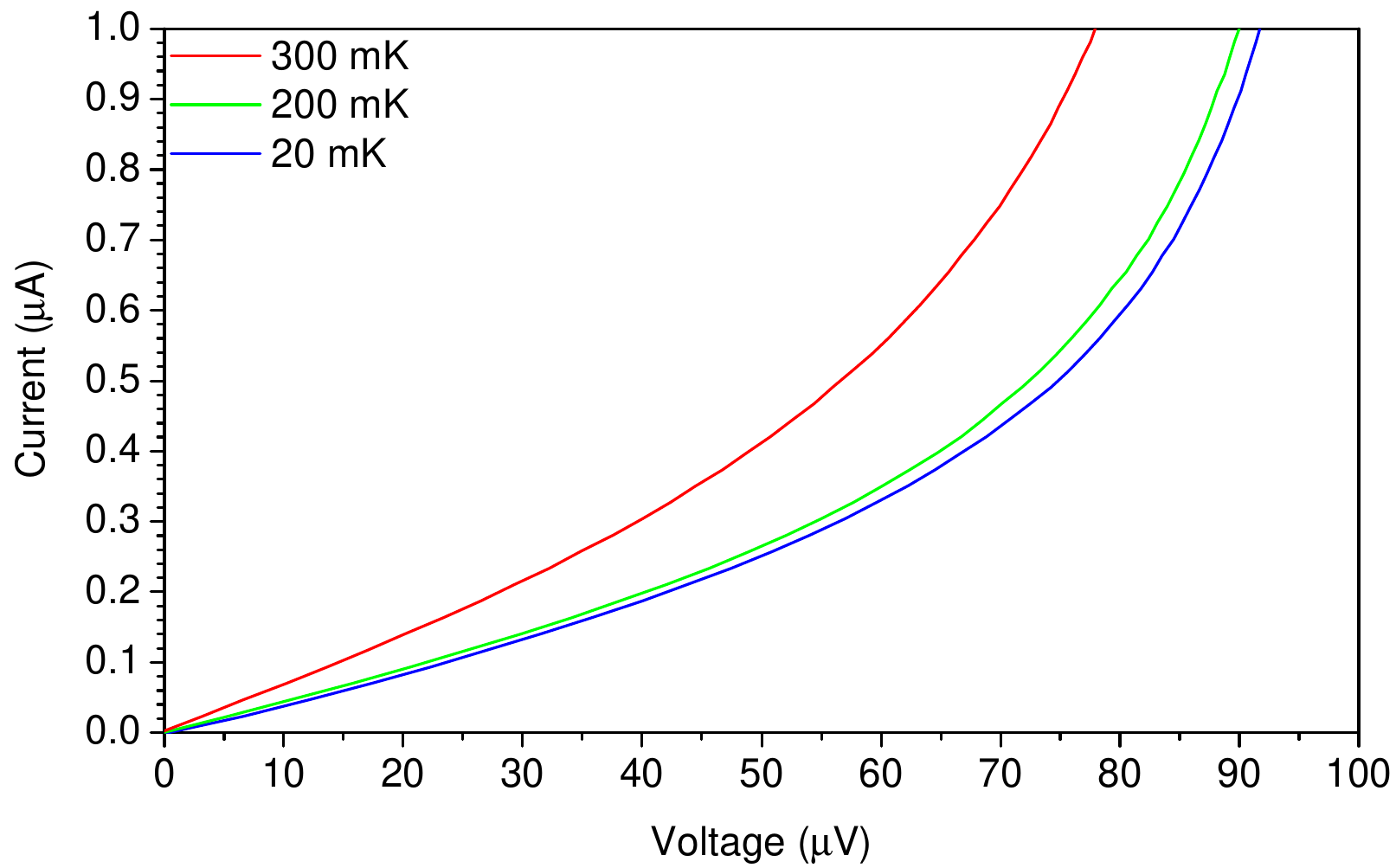}    
    \caption{Current-voltage characteristics of the sample with aluminum antenna measured at temperatures of 20-300 mK.}
    \label{fig:sample_Al_IV}
\end{figure}

Current-voltage characteristics of the sample with aluminum made antenna for various temperatures are shown in Figure \ref{fig:sample_Al_IV}.

If one looks at the presented current-voltage characteristics, it might be seen that the current is much higher, and the voltage is lower than for previous sample with the normal metal antenna. It can be explained by parasitic tunnel junctions that occur because of the use of all-aluminum technology. Due to this reason, the fitting procedure is impossible to be performed, because the SIN junction parameters are masked.

\begin{figure}
 \includegraphics[width=\linewidth, keepaspectratio]{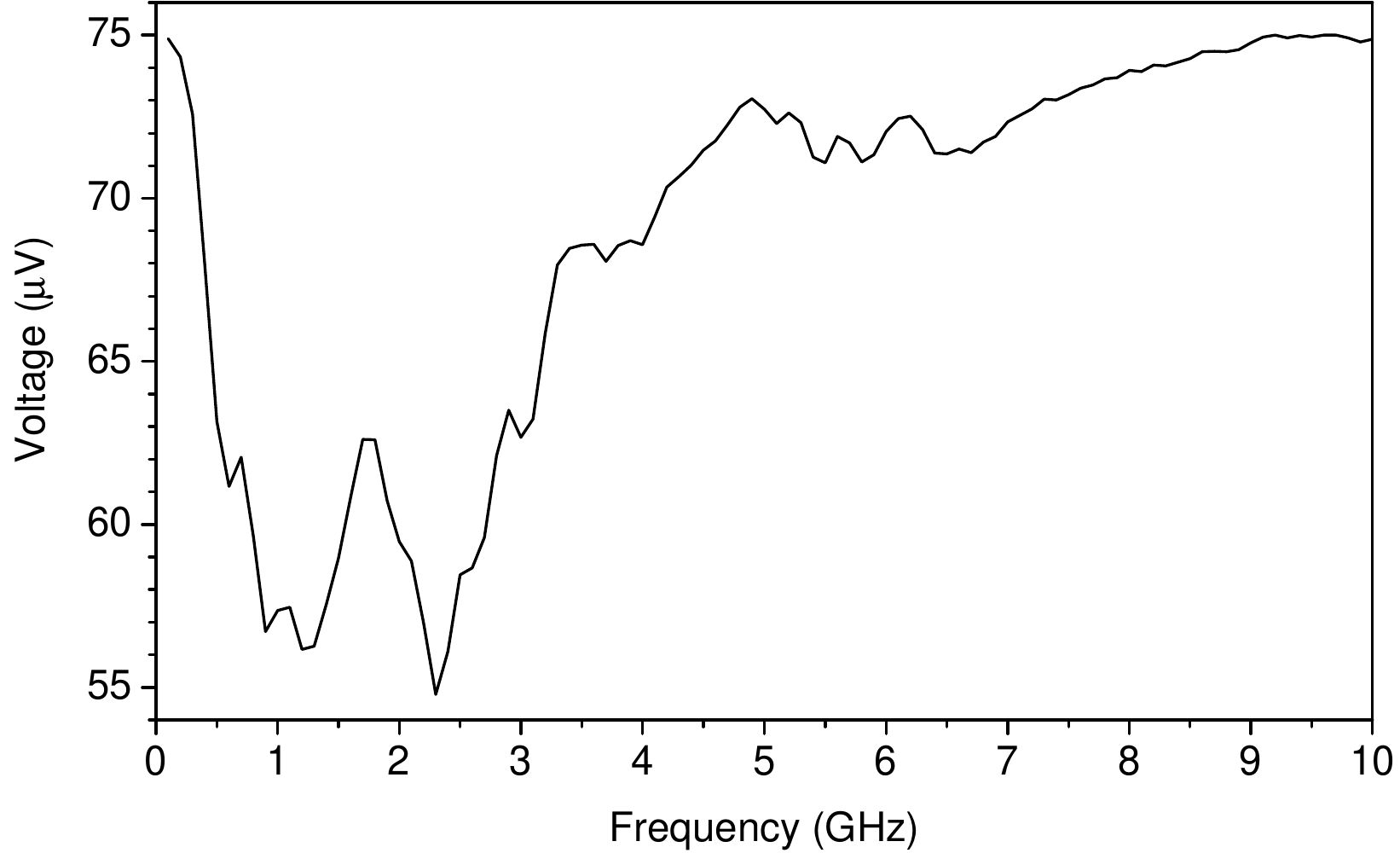}    
    \caption{AFC of the sample with aluminum antenna measured at the temperature of 300 mK.}
    \label{fig:sample_Al_AFC}
\end{figure}

The AFC of the sample is shown in Fig. \ref{fig:sample_Al_AFC}. One can see that, while the response shape is similar to the previous sample with the normal metal antenna, see Fig. \ref{fig:sample_response}, the central frequency is significantly shifted to lower frequencies of 0.5--3 GHz. The shift of the response frequency range can be explained by the contribution of kinetic inductance of the superconducting antenna.

\section{Conclusions}

The samples realized with a novel concept of a CEB integrated into a coplanar antenna were fabricated and tested. The measurements show the high characteristics of this design. The modification, introducing hafnium sublayer has shown its efficiency in improving the quality of a tunnel barrier, thus decreasing leakage current. 

The sample with coplanar antenna and DC lines made of Ti/Au/Pd trilayer has shown a double response peak at the frequencies of 7.7 GHz and 8.8 GHz, and also a single peak at 14 GHz. Due to the efficient electron cooling from 300 mK to 71 mK, the minimal NEP value reaches an impressive value of $6 \, \rm{aW/\sqrt{Hz}}$, unattainable before for other types of detectors at the same 300 mK plate temperature.  This allows operation in space in $^3$He sorption cryostats without necessity to use dilution refrigerators.

Another measured sample with aluminum antenna has shown efficient response in the frequency range 0.5--3 GHz. While the antenna design is the same as for the other sample with Ti/Au/Pd antenna, the observed frequency shift is explained by kinetic inductance of superconductive aluminum, which should be taken into account in future designs.

\ack

This work was supported by Russian Science Foundation, project 21-79-20227.

Lithography and TEM images acquisition were performed using the facilities of the Common Research Centre “Physics and technology of micro and nanostructures” of IPM RAS. Electron beam deposition was performed using the facilities of the Laboratory of Superconducting Nanoelectronics of NNSTU.

The samples with aluminum antennas were fabricated in the Institute of Nanotechnologies of Microelectronics of RAS.

The authors declare no conflict of interests.

\section{Data availability statement}
The data that support the findings of this study
are available from A.L. Pankratov upon reasonable
request.

\end{document}